\documentclass[conference]{IEEEtran}
\usepackage{ifpdf}

\usepackage{cite}

\ifCLASSINFOpdf
\else
\fi
\usepackage{amssymb}
\usepackage[cmex10]{amsmath}
\usepackage{algorithmic}

\usepackage{array}

\usepackage{mdwmath}
\usepackage{mdwtab}

\usepackage{eqparbox}

\usepackage{graphicx}
\usepackage{epstopdf}
\usepackage[tight,footnotesize]{subfigure}

\usepackage{fixltx2e}

\usepackage{stfloats}

\usepackage{url}

\IEEEoverridecommandlockouts 
\begin{document}
\newcommand{\bpi}{\mbox{\boldmath{ $\pi $}}}
\newcommand{\bx}{{\mbox{\boldmath{$x$}}}}
\title{The Trajectory PHD Filter for Jump Markov System Models and Its Gaussian Mixture Implementation}

\author{Boxiang~Zhang,
	Wei~Yi,~\IEEEmembership{Member,~IEEE}
	
}


%


\maketitle

\begin{abstract}
The trajectory probability hypothesis density filter (TPHD) is capable of producing trajectory estimates in first principle without adding labels or tags. In this paper, we propose a new TPHD filter referred as MM-TPHD for jump Markov system (JMS) model that the highly dynamic targets movement switches between multiple models. Firstly, we extend the concept of JMS to set of trajectories and derive the TPHD recursion for the proposed JMS model. Then, we develop the linear Gaussian Mixture (LGM) implementation of MM-TPHD recursion and also consider the $L$-scan computationally efficient implementations. Finally, in a challgenging multiple maneuvering targets tracking scenario, the simulation results demonstrate the performance of the proposed algorithm.
\end{abstract}


%
\IEEEpeerreviewmaketitle
\section{Introduction}
In most radar applications,, such as vehicle radar and shipborne radar. The multiple maneuvering targets tracking~\cite{mmjm} involves jointly estimating the time-varying number of targets and their states from a set of observations in the presence of target maneuver uncertainty, data association uncertainty, detection uncertainty, noise and clutter. Hence, the research topic is highly challenging both in theoretical derivation and algorithm implementation.

The jump Markov system (JMS) or multiple models approach, in which the target state is augmented with an additional motion model label and the model evolves with time according to a finite state Markov chain~\cite{imma}, is a popular approach for single maneuvering targets tracking~\cite{mmjm,imma,pfjms}. In order to track multiple maneuvering targets, besides combining traditional data association algorithms such as joint probabilistic data association (JPDA)~\cite{cmjpda,mjpda} or multiple hypothesis tracking (MHT)~\cite{mmht} with JMS models, the RFS approach~\cite{msmm,mome} is also an attractive tool. The RFS approach has been adopted to formulate multiple models extensions of PHD~\cite{mphd}, GM-PHD~\cite{mmphd}, CPHD~\cite{mmcphd,mcphd}, multi-Bernoulli~\cite{mmmb}, LMB~\cite{mmlmb} and GLMB~\cite{mmglmb,mmvo} filters.

Recently, the principle approach of forming trajectories has become more and more interesting. To date, two major solution paradigms have been emerged. These are, multi-target state sequence posterior~\cite{msglmb} and \textit{set of trajectories} $/$ \textit{trajectory random finite set (RFS)}~\cite{tcphd,mttst,pmbmst}. In the formulation of multi-target state sequence posterior, the multi-scan generalized labeled multi-Bernoulli (GLMB) filter~\cite{msglmb} shows the excellent multi-target and multi-target tracking performance comparing with the GLMB filter~\cite{conj} who is an analytic solution to the multi-target Bayes filter. By contrast, the trajectory RFS approach is computationally efficient, although its trajectory tracking performance is not better than the former.  

Considering the trajectory RFS approach, the TPHD filter~\cite{tcphd} is capable to estimate the trajectories of the alive targets by propagating a Poisson cluster multi-trajectory density through the filtering recursion using KLD minimisations. The closed-form solution for single linear Gaussian model is presented in~\cite{tcphd}. However, the single model is powerless for multiple maneuvering target system as it obeys jump Markov system (JMS) model that the highly dynamic targets movement switches between multiple models.

In this paper, we generalize the concept of JMS to the trajectory RFS formulation of multiple maneuvering targets. Combined with the JMS model, we present a new TPHD filter to track the trajectories accommodating births, deaths and switching dynamics at each time step, named MM-TPHD filter. Then, the MM-TPHD recursion is derived and we develop the LGM implementation in which case we can implement the MM-TPHD filter in analytic closed-form. The $L$-scan approximation of the LGM implementation is also considered to deal with the computational infeasibility caused by the case that the length of trajectory increases with time. In addition, simulation results verify the accurate trajectory tracking performance of the MM-TPHD filter in multiple maneuvering targets tracking scenario.
\section{Background}
In this section, we briefly review the trajectory RFS, multi-trajectory Bayes recursion and the TPHD filter~\cite{tcphd}.
\subsection{Trajectory RFS}

According to the trajectory state model proposed in~\cite{tcphd}, a single trajectory kinematic state is represented as a variable $X=\left( \beta ,{{x}^{1:l}} \right)$, where $\beta $ is the birth time of the trajectory, $l$ is its length and ${{x}^{1:l}}\text{=}\left( {{x}^{1}},\cdots ,{{x}^{l}} \right)$ denotes the continuous states sequence of the trajectory, ${{x}^{k}}\in {{\mathbb{R}}^{{{n}_{x}}}}$ is the single target kinematic state. Then, we denote the trajectory state space at time $k$ as follows
\begin{equation}\label{eq1}
{{\mathcal{T}}_{k}}={{\uplus }_{(\beta ,l)\in {{J}_{k}}}}\left\{ \beta  \right\}\times {{\mathbb{R}}^{l{{n}_{x}}}},
\end{equation}
where ${{J}_{k}}=\left\{ \left( \beta ,l \right):0\le \beta \le k,1\le l\le k-\beta +1 \right\}$, $\uplus $ denotes disjoint set union.

Given $X\in {{\mathcal{T}}_{k}}$, the trajectory state density is
\begin{equation}\label{eq2}
p\left( X \right)=p\left( {{x}^{1:l}}|\left( \beta ,l \right) \right)P\left( \beta ,l \right),
\end{equation}
where $\left( \beta ,l \right)\in {{J}_{k}}$. And the integral of trajectory state density is expressed as
\begin{equation}\label{eq3}
\int{p\left( X \right)dX}=\sum\limits_{\left( \beta ,l \right)\in {{J}_{k}}}{P\left( \beta ,l \right)\int{p\left( {{x}^{1:l}}|\left( \beta ,l \right) \right)d}{{x}^{1:l}}}.
\end{equation}

Similar to the set of targets, we define the set of trajectories at time $k$ as ${{\textbf{X}}_{k}}\in \mathcal{F}({{\mathcal{T}}_{k}})$ 
\begin{equation}\label{eq4}
{{\textbf{X}}_{k}}\text{=}\left\{ X=\left( \beta ,{{x}^{1:l}} \right)\in {{\mathcal{T}}_{k}} \right\},
\end{equation}
then we denotes $\pi \left( \textbf{X} \right)$ as the multi-trajectory density on a set of trajectories $\textbf{X}$, thus the set integral~\cite{msmm} of $\textbf{X}$ is defined by
\begin{equation}\label{eq5}
\int{\pi \left( \textbf{X} \right)\delta }\textbf{X}=\pi (\varnothing )+\sum\limits_{n=1}^{\infty }{\frac{1}{n!}\int{\cdots }\int{\pi (\left\{ {{X}_{1}},\cdots ,{{X}_{n}} \right\})}d{{X}_{1:n}}},
\end{equation}
and its cardinality distribution $\rho \left( \cdot  \right)$ is
\begin{equation}\label{eq6}
\rho \left( n \right)=\frac{1}{n!}\int{\cdots }\int{\pi (\left\{ {{X}_{1}},\cdots ,{{X}_{n}} \right\})}d{{X}_{1:n}}.
\end{equation}

\textit{Poisson Trajectory RFS}: A trajectory RFS $\textbf{X}$ is referred as Poisson if its cardinality $\left| \textbf{X }\right|$ is Poisson distributed with mean ${{\lambda }_{v}}$, and the elements of $\textbf{X}$ are independently and identically distributed (i.i.d.) according to the probability density $\overset{\scriptscriptstyle\smile}{v}\left( \cdot  \right)$.

The probability density of the Poisson trajectory RFS is given by~\cite{tcphd}
\begin{equation}\label{eq7}
\pi (\left\{ {{X}_{1}},\cdots {{X}_{n}} \right\})={{e}^{-{{\lambda }_{v}}}}\lambda _{v}^{n}\prod\limits_{i=1}^{n}{\overset{\scriptscriptstyle\smile}{v}\left( {{X}_{i}} \right)}.
\end{equation}
\subsection{Bayesian Multi-trajectory Recursion }
Conditional on the multi-trajectory posterior density ${{\pi }_{k-1}}\left( {{\textbf{X}}_{k-1}} \right)$ at time $k-1$, the multi-trajectory posterior density ${{\pi }_{k}}\left( {{\textbf{\textbf{X}}}_{k}} \right)$ is calculated via the prediction and update as follows~\cite{mttst}
\begin{align}\label{eq8}
{{\pi }_{k|k-1}}\left( {{\textbf{X}}_{k}} \right)&=\int{{{f}_{k|k-1}}}\left( {{\textbf{X}}_{k}}|\textbf{X}{}_{k-1} \right){{\pi }_{k-1}}\left( {{\textbf{X}}_{k-1}} \right)\delta {{\textbf{X}}_{k-1}},\\
{{\pi }_{k}}\left( {{\textbf{X}}_{k}} \right)&=\frac{{{g}_{k}}\left( {{\textbf{z}}_{k}}|{{\textbf{X}}_{k}} \right){{\pi }_{k|k-1}}\left( {{\textbf{X}}_{k}} \right)}{{{h}_{k}}\left( {{\textbf{z}}_{k}} \right)},
\end{align}
where ${{\pi }_{k|k-1}}\left( {{\textbf{X}}_{k}} \right)$ is the predicted multi-trajectory density at time $k$,  ${{\textbf{z}}_{k}}$ is multi-target observation at time $k$, ${{f}_{k|k\text{-}1}}(\cdot |\cdot )$ is the multi-trajectory transition kernel and ${{g}_{k}}\left( \cdot |\cdot  \right)$ is the density of the measurements given the current RFS of trajectory. ${{h}_{k}}\left( {{\textbf{z}}_{k}} \right)$ is the normalizing constant.

Suppose ${{\textbf{\textbf{x}}}_{k}}$ denotes the corresponding multi-target state set for the set of trajectories ${{\textbf{X}}_{k}}$ at time $k$, as a result, we can reach that ${{g}_{k}}\left( {{\textbf{z}}_{k}}|{{\textbf{X}}_{k}} \right)\text{=}{{l}_{k}}\left( {{\textbf{z}}_{k}}|{{\textbf{x}}_{k}} \right)$ intuitively, where ${{l}_{k}}\left( \cdot |\cdot  \right)$ is the multi-target likelihood function at time $k$.
\subsection{TPHD Filter}
The PHD that represents the first-order statistical moment of multi-trajectory density $\pi \left( \textbf{X} \right)$, is defined by~\cite{tcphd}
\begin{equation}\label{eq10}
{{D}_{\pi }}\left( X \right)=\int{\pi \left( \left\{ X \right\}\cup \textbf{X} \right)}\delta \textbf{X},
\end{equation}
and the Poisson multi-trajectory density can be characterized by its PHD ${{D}_{v}}\left( X \right)={{\lambda }_{v}}\overset{\scriptscriptstyle\smile}{v}\left( X \right)$~\cite{msmm}. 

Instead of propagating the best Poisson approximation for the multi-trajectory density straightforwardly, the TPHD filter recursively propagates the posterior intensity (PHD) of the Poisson multi-trajectory density, in the sense of minimizing the Kullback-Leibler divergence (KLD).

\textit{\textbf{Prediction}}: In the prediction step, the following assumptions are taken:
\begin{itemize}
	\item [P1] 
	Given the current multi-target state $\textbf{x}$, each target $x\in \textbf{x}$ either continues to survive with probability ${{P}_{S}}\left( x \right)$ and moves to a new state with transition density $t\left( \cdot |x \right)$ or dies with probability $\text{1-}{{P}_{S}}\left( x \right)$.       
	\item [P2]
    The multi-target set at next time is the union of the surviving targets last time and the current new targets that are born independently with a Poisson multi-target density ${{\gamma }_{\tau }}\left( \cdot  \right)$  
	\item [P3]
    The multi-target RFS at time $k-1$ is Poisson. 
\end{itemize}

Note that the subindex $\tau $ represents the density of target RFS. Under Assumptions P1-P3, given the posterior PHD ${{D}_{k-1}}\left( X \right)$ at time $k-1$, the predicted PHD ${{D}_{k|k-1}}\left( X \right)$ at time $k$ as follows
\begin{equation}\label{eq11}
{{D}_{k|k-1}}\left( X \right)\text{=}{{D}_{\gamma ,k}}\left( X \right)+{{D}_{\zeta ,k}}\left( X \right),
\end{equation}
where

\begin{align}\label{eq12}
{{D}_{\gamma ,k}}\left( \beta ,{{x}^{1:l}} \right)&={{D}_{{{\gamma }_{\tau }}}}\left( {{x}^{1}} \right){{1}_{\left\{ k \right\}}}\left( \beta  \right),\\
\label{eq13}
 {{D}_{\zeta ,k}}\left( \beta ,{{x}^{1:l}} \right)&={{P}_{S}}\left( {{x}^{l}} \right)t\left( {{x}^{l}}|{{x}^{l-1}} \right){{D}_{k-1}}\left( \beta ,{{x}^{1:l-1}} \right),
\end{align}
where ${{\mathbb{N}}_{k-1}}=\left\{ 1,\cdots ,k-1 \right\}$ and ${{1}_{\text{Y}}}(x)$ is the inclusion function.

As \eqref{eq11}, the predicted PHD contains the PHD of the newborn trajectories and the PHD of the surviving trajectories. The termination time of a trajectory $X=\left( \beta ,{{x}^{1:l}} \right)$ is $\beta +l-1$, thus the predicted PHD is zero if $\beta +l-1\ne k$ that indicates the trajectory is dead, as this paper only considers the alive trajectories.

\textit{\textbf{Update}}: In the update step, the following assumptions are taken:
\begin{itemize}
	\item [U1] 
	Given the current multi-target state $\textbf{x}$, each target $x\in \textbf{x}$ is either detected with probability ${{P}_{D}}\left( x \right)$ and generates a measurement $z$ with likelihood $l(\cdot |x)$ or missed with probability $\text{1-}{{P}_{D}}\left( x \right)$.       
	\item [U2]
	The multi-target observation $\textbf{z}$ is the superposition of the observations from detected targets and Poisson clutter with intensity $\kappa \left( \cdot  \right)$. 
	\item [U3]
	The predicted multi-trajectory RFS at time $k$ is Poisson. 
\end{itemize}

Under Assumptions U1-U3, given the predicted PHD ${{D}_{k|k-1}}\left( X \right)$, the updated PHD ${{D}_{k}}\left( X \right)$ at time $k$ is
\begin{equation}\label{eq14}
\begin{aligned}
& {{D}_{k}}\left( X \right)\text{=}{{D}_{k}}\left( \beta ,{{x}^{1:l}} \right)={{D}_{k|k-1}}\left( \beta ,{{x}^{1:l}} \right)\times  \\ 
& \left( 1-{{P}_{D}}\left( {{x}^{l}} \right)+\sum\limits_{z\in {{\textbf{z}}_{k}}}{\frac{{{P}_{D}}\left( {{x}^{l}} \right)l\text{(}z|{{x}^{l}}\text{)}}{\kappa \left( z \right)+\int{{{P}_{D}}\left( {{x}^{l}} \right)l\text{(}z|{{x}^{l}}\text{)}{{D}_{k|k-1}}\left( {{x}^{l}} \right)d{{x}^{l}}}}} \right) \\  
\end{aligned},
\end{equation}
where $l$ = $k-\beta +1$ or ${{D}_{k}}\left( X \right)=0$, otherwise, and ${{D}_{k|k-1}}\left( {{x}^{l}} \right)$ denotes the PHD of the targets at time $k$, which is defined as ~\cite{tcphd}
\begin{equation}\label{eq15}
{{D}_{k|k-1}}\left( {{x}^{l}} \right)\text{=}\sum\limits_{\beta =1}^{k}{\int{{{D}_{k|k-1}}\left( \beta ,{{x}^{1:l}} \right)d{{x}^{1:l-1}}}}.
\end{equation}

Analogously to the PHD update~\cite{phd,mome}, the TPHD update also only concerns the associations between single target and all measurements. We only present the result of alive trajectories in this paper and the prediction, update are proven in~\cite{tcphd} for a more general case in which all trajectories including dead trajectories are considered. 
\section{JMS TPHD Filter}
This section presents a new TPHD filter referred as MM-TPHD that can accommodate maneuvering targets that the highly dynamic targets motion switches between multiple models. The JMS model of trajectory RFS is described in Section III-A. We derive the relevant TPHD recursion in Section III-B. Then, the linear Gaussian mixture implementation and the L-scan computationally efficient implementations is developed in Section III-C, III-D, respectively.
 
To describe the motion of maneuvering targets, an additional variable $o\in \mathbb{O}$ that denotes the label of motion model or the mode is adopted, where $\mathbb{O}$ represents the discrete space of all possible modes. Thus, the single trajectory state is defined as an augmented vector $\bar{X}$ = $\left( X,O \right)$ = $\left( \beta ,{{x}^{1:l}},{o} \right)$ $\in$ $ \mathcal{T}\times {{\mathbb{O}}}$, where the mode $o$ means the motion model of the trajectory at current time. The augmented trajectory RFS is denoted as
\begin{equation}\label{eq16}
\bar{\textbf{X}}\text{=}\left\{ \bar{X}=\left( \beta ,{{x}^{1:l}},{{o}} \right)\in \mathcal{T}\times {{\mathbb{O}}} \right\}.
\end{equation}


\subsection{Jump Markov System}
A JMS can be expressed as a set of parameterized state space models whose parameters change with time according to finite state Markov chain. Let $\upsilon \left( o|{o}' \right)$ denotes the model switch probability from motion model ${o}'$ to motion model $o$. Then, the sum of the switch probabilities of all possible motion model given motion model adds up to 1, i.e., $\sum\limits_{o\in \mathbb{O}}{\upsilon \left( o|{o}' \right)}\text{=}1$. 

In some applications, the motion switch is independent of the state transition. Thus, the transition probability of augmented single trajectory state is denoted as
\begin{equation}\label{eq17}
f\left( \bar{X}|{\bar{X}}' \right)=f\left( X,o|{X}',{o}' \right)=f\left( X|{X}',{o}' \right)\upsilon \left( o|{o}' \right),
\end{equation}
and the measurement likelihood function is generally independent of motion model, therefore, we express the trajectory-measurement likelihood function as 
\begin{equation}\label{eq18}
g\left( z|\bar{X} \right)=g\left( z|X,o \right)=g\left( z|X \right).
\end{equation}

In the TPHD filter, what really works are single target transition function $t\left( \cdot |x \right)$ and single target-measurement likelihood function $l(\cdot |x)$, shown as \eqref{eq13},\eqref{eq14}. Consequently, we define the augmented single-target state as $\bar{x}=\left( x,o \right)$.The transition function and measurement likelihood function for the augmented single target state can be expressed as
\begin{align}\label{eq19}
t\left( \bar{x}|{\bar{x}}' \right)&=t\left( x,o|{x}',{o}' \right)=t\left( x|{x}',{o}' \right)\upsilon \left( o|{o}' \right),\\
\label{eq20}
l(z|\bar{x})&=l(z|x,o)=l(z|x).
\end{align}
\subsection{TPHD Filter for JMS Models}
Combined with JMS model, we express the PHD of the augmented trajectory RFS $\bar{\textbf{X}}$ as $D\left( {\bar{X}} \right)$=$D\left( \beta ,{{x}^{1:l}},o \right)$. The recursive details of the MM-TPHD filter as follows. 

\textbf{\textit{Prediction:}} In the MM-TPHD prediction step, the assumptions P1-P3 are still adopted, but we need to replace kinematic state with augmented state.

\textit{Proposition 1:} Given the posterior PHD ${{D}_{k-1}}\left( {\bar{X}} \right)$ at time $k-1$, the predicted PHD ${{D}_{k|k-1}}\left( {\bar{X}} \right)$ at time $k$ is given by
\begin{equation}\label{eq21}
{{D}_{k|k-1}}\left( {\bar{X}} \right)\text{=}{{D}_{\gamma ,k}}\left( {\bar{X}} \right)+{{D}_{\zeta ,k}}\left( {\bar{X}} \right),
\end{equation}
where
\begin{align}\label{eq22}
{{D}_{\gamma ,k}}\left( {\bar{X}} \right)&={{1}_{\left\{ k \right\}}}\left( \beta  \right){{D}_{{{\gamma }_{\tau }}}}\left( {{x}^{1}},{{o}^{1}} \right),\\
\begin{split}\label{eq23}
{{D}_{\zeta ,k}}\left( {\bar{X}} \right)&={{1}_{{{\mathbb{N}}_{k-1}}}}\left( \beta  \right){{P}_{S}}\left( {{x}^{l}},{{o}^{l}} \right){{D}_{\zeta}}\\
&\times\sum\limits_{{{o}^{l-1}}\in \mathbb{O}}{t\left( {{x}^{l}}|{{x}^{l-1}},{{o}^{l}} \right)\upsilon\left( {{o}^{l}}|{{o}^{l-1}} \right){{D}_{k-1}}\left( \beta ,{{x}^{1:l-1}},{{o}^{l-1}}\right)}.
\end{split}
\end{align}

\textbf{\textit{Update:} }In the MM-TPHD update step, the assumptions U1-U3 are also taken. As mentioned above, the measurement likelihood function is generally independent of mode. 

\textit{Proposition 2:} Given the predicted PHD ${{D}_{k|k-1}}\left( {\bar{X}} \right)$ at time $k$, the posterior PHD ${{D}_{k}}\left( {\bar{X}} \right)$ at time $k$ is given by
\begin{equation}\label{eq24}
{{D}_{k}}\left( {\bar{X}} \right)={{D}_{mis,k}}\left( {\bar{X}} \right)+{{D}_{det ,k}}\left( {\bar{X}} \right),
\end{equation}
where
\begin{align}\label{eq25}
{{D}_{mis,k}}\left( {\bar{X}} \right)&=\left( 1-{{P}_{D}}\left( {{x}^{l}},{{o}^{l}} \right) \right){{D}_{k|k-1}}\left( \beta ,{{x}^{1:l}},{{o}^{l}} \right),\\
{{D}_{det ,k}}\left( {\bar{X}} \right)&={{D}_{k|k-1}}\left( \beta ,{{x}^{1:l}},{{o}^{l}} \right)\sum\limits_{z\in {{\textbf{z}}_{k}}}{\frac{{{P}_{D}}\left( {{x}^{l}},{{o}^{l}} \right)l\left( z|{{x}^{l}},{{o}^{l}} \right)}{\kappa \left( z \right)+\varepsilon }},\\
\varepsilon &=\int{\sum\limits_{{{o}^{l}}\in \mathbb{O}}{{{P}_{D}}\left( {{x}^{l}},{{o}^{l}} \right)l\left( z|{{x}^{l}},{{o}^{l}} \right){{D}_{k|k-1}}\left( {{x}^{l}},{{o}^{l}} \right)}d{{x}^{l}}},\\
{{D}_{k|k-1}}\left( {{x}^{l}},{{o}^{l}} \right)&=\sum\limits_{\beta =1}^{k}{\int{{{D}_{k|k-1}}\left( \beta ,{{x}^{1:l}},{{o}^{l}} \right)d{{x}^{1:l-1}}}}.
\end{align}

Proposition 1 and 2 show how the trajectory posterior intensity is propagated in time on the JMS multi-target model. The analytic solution for the MM-TPHD filter based on Gaussian mixture (GM) approximate of the PHD is presented in next subsection.

\subsection{LGM Implementation}
The LGM implementation of MM-TPHD is presented in this subsection. We use the notation
\begin{equation}\label{eq29}
\begin{aligned}
\mathcal{N}\left( \beta ,{{x}^{1:l}},o;{{\beta }_{k}},{{m}_{k}},{{U}_{k}} \right)&={{\delta }_{{{\beta }_{k}}\left( o \right)}}\left( \beta  \right){{\delta }_{{{l}_{k}}\left( o \right)}}\left( l \right)\\
&\times\mathcal{N}\left( {{x}^{1:l}};{{m}_{k}}\left( o \right),{{U}_{k}}\left( o \right) \right), 
\end{aligned}
\end{equation}
 where ${{l}_{k}}\left( o \right)$=$\dim\left( {{m}_{k}}\left( o \right)/{{n}_{x}} \right)$. \eqref{eq29} represents a single trajectory Gaussian density with mode $o$, start time ${{\beta }_{k}}\left( o \right)$, length ${{l}_{k}}\left( o \right)$, mean ${{m}_{k}}\left( o \right)\in {{\mathbb{R}}^{{{l}_{k}}\left( o \right){{n}_{x}}}}$ and covariance ${{U}_{k}}\left( o \right)\in {{\mathbb{R}}^{{{l}_{k}}\left( o \right){{n}_{x}}\times {{l}_{k}}\left( o \right){{n}_{x}}}}$.

In addition, we take some assumptions as follows
\begin{itemize}
	\item [A1] 
	The survival probability and detection probability are constants, i.e., ${{P}_{S}}\left( x,o \right)\text{=}{{P}_{S}}$, ${{P}_{D}}\left( x,o \right)\text{=}{{P}_{D}}$.      
	\item [A2]
	Both the transition density and measurement likelihood are linear Gaussian,
	 \begin{align}\label{eq30}
	  t\left( {{x}^{l}}|{{x}^{l-1}},{{o}^{l}} \right)&=\mathcal{N}\left( {{x}^{l}};F\left( o \right){{x}^{l-1}},Q\left( o \right) \right), \\ 
	  \label{eq31}
	  l\left( z|{{x}^{l}},{{o}^{l}} \right)&=l\left( z|{{x}^{l}} \right)=\mathcal{N}\left( z;H{{x}^{l}},R \right),  
	 \end{align}
	 where $F\in {{\mathbb{R}}^{{{n}_{x}}\times {{n}_{x}}}}$ is the single target transition matrix, $Q\in {{\mathbb{R}}^{{{n}_{x}}\times {{n}_{x}}}}$ is the covariance matrix of single target process noise and $F$, $Q$ depend on the mode of target. $H\in {{\mathbb{R}}^{{{n}_{z}}\times {{n}_{x}}}}$ is the single measurement matrix and $R\in {{\mathbb{R}}^{{{n}_{z}}\times {{n}_{z}}}}$ is the covariance matrix of single measurement noise.
	\item [A3]
	The PHD of the birth density ${{\gamma }_{\tau }}\left( \cdot  \right)$ at time $k$ is a Gaussian mixture
	 \begin{equation}\label{eq32}
	 \begin{aligned}
	 &{{D}_{\gamma ,k}}\left( {\bar{X}} \right)=\\
	 &\sum\limits_{j=1}^{{{J}_{\gamma ,k}}\left( {{o}^{k}} \right)}{\omega _{\gamma ,k}^{j}\left( {{o}^{k}} \right)\mathcal{N}\left( \bar{X};k,m_{\gamma ,k}^{j}\left( {{o}^{k}} \right),U_{\gamma ,k}^{j}\left( {{o}^{k}} \right) \right)},
	 \end{aligned} 
	 \end{equation}
	 where ${{J}_{\gamma ,k}}\in \mathbb{N}$ is the number of Gaussian components, $\omega _{\gamma ,k}^{j}$ is the weight of the $j$th component, $m_{\gamma ,k}^{j}$ and $U_{\gamma ,k}^{j}$ are its mean and covariance matrix, respectively.
\end{itemize}

Note that the models provided by A1-A3 are time-varying but time index is omitted for notational convenience. Under Assumptions A1-A3, P1-P3 and U1-U3, we can implement the LGM-MM-TPHD in analytic closed-form as follows.

\textit{\textit{Proposition 3 (prediction):}} If the PHD ${{D}_{k-1}}\left( {\bar{X}} \right)$ of the augmented trajectory RFS at time $k-1$ has the form
\begin{equation}\label{eq33}
\begin{aligned}
{{D}_{k-1}}\left( {\bar{X}} \right)&=\sum\limits_{j=1}^{{{J}_{k\text{-}1}}\left( {{o}^{k-1}} \right)}{\omega _{k-1}^{j}\left( {{o}^{k-1}} \right)}\\
	&\times\mathcal{N}\left( \bar{X};\beta _{k-1}^{j}\left( {{o}^{k-1}} \right),m_{k-1}^{j}\left( {{o}^{k-1}} \right),U_{k-1}^{j}\left( {{o}^{k-1}} \right) \right),\\ 
\end{aligned}
\end{equation}
where $\beta _{k-1}^{j}\left( {{o}^{k-1}} \right)+{{l}_{k-1}}-1=k-1$ as we just consider the alive trajectories. Then, the predicted PHD ${{D}_{k|k-1}}\left( {\bar{X}} \right)$ at time $k$ is
\begin{equation}\label{eq34}
{{D}_{k|k-1}}\left( {\bar{X}} \right)\text{=}{{D}_{\gamma ,k}}\left( {\bar{X}} \right)+{{D}_{\zeta ,k}}\left( {\bar{X}} \right),
\end{equation}
where
\begin{align}
\begin{split}\label{eq35}
{{D}_{\zeta ,k}}\left( {\bar{X}} \right)&=\sum\limits_{j=1}^{{{J}_{k\text{-}1}}\left( {{o}^{k-1}} \right)}{\sum\limits_{{{o}^{k-1}}\in \mathbb{O}}{\omega _{k|k-1}^{j}\left( {{o}^{k}} \right)}}\\
&\times\mathcal{N}\left( \bar{X};\beta _{k-1}^{j}\left( {{o}^{k-1}} \right),m_{k|k-1}^{j}\left( {{o}^{k}} \right),U_{k|k-1}^{j}\left( {{o}^{k}} \right) \right), 
\end{split}\\
\omega _{k|k-1}^{j}\left( {{o}^{k}} \right)&={{P}_{S}}\upsilon \left( {{o}^{k}}|{{o}^{k-1}} \right)\omega _{k-1}^{j}\left( {{o}^{k-1}} \right),\\
m_{k|k-1}^{j}\left( {{o}^{k}} \right)&=\left[ \begin{matrix}
m_{k-1}^{j}\left( {{o}^{k-1}} \right)  \\
F\left( {{o}^{k}} \right)m_{k-1}^{j,\left[ k-1 \right]}\left( {{o}^{k-1}} \right)  \\
\end{matrix} \right],
\end{align}
\begin{align}
U_{k|k-1}^{j}\left( {{o}^{k}} \right)&=\left[ \begin{matrix}
U_{k-1}^{j}\left( {{o}^{k-1}} \right) & {{U}_{1}}  \\
U_{1}^{\top} & {{U}_{2}}+Q\left( {{o}^{k}} \right)  \\
\end{matrix} \right],\\
{{U}_{1}}&=U_{k-1}^{j,\left[ :,k-1 \right]}\left( {{o}^{k-1}} \right)F{{\left( {{o}^{k}} \right)}^{\top}},\\
{{U}_{2}}&=F\left( {{o}^{k}} \right)U_{k-1}^{j,\left[ k-1 \right]}\left( {{o}^{k-1}} \right)F{{\left({{o}^{k}} \right)}^{\top}},
\end{align}
where $m_{k}^{j,\left[ a \right]}$ and $U_{k}^{j,\left[ a \right]}$ denote the parts of the mean vector and the covariance matrix of the $j$th component for time step $a$~\cite{pmbmst}, and $U_{k}^{j,\left[ a:b,c:d \right]}$ denotes the part the covariance matrix with rows for time steps $a$ to $b$ and columns for time steps $c$ to $d$.

\textit{Proposition 4 (update):} If the predicted PHD ${{D}_{k|k-1}}\left( {\bar{X}} \right)$ at time $k$ has the form
\begin{equation}\label{eq57}
\begin{aligned}
{{D}_{k|k-1}}\left( {\bar{X}} \right)&=\sum\limits_{j=1}^{{{J}_{k|k-1}}\left( {{o}^{k}} \right)}{\omega _{k|k-1}^{j}\left( {{o}^{k}} \right)}\\
&\times\mathcal{N}\left( \bar{X};\beta _{k|k-1}^{j}\left( {{o}^{k}} \right),m_{k|k-1}^{j}\left( {{o}^{k}} \right),U_{k|k-1}^{j}\left( {{o}^{k}} \right) \right).\\ 
\end{aligned}
\end{equation}

Then, the posterior PHD ${{D}_{k}}\left( {\bar{X}} \right)$ at time $k$ is  
\begin{equation}\label{eq41}
{{D}_{k}}\left( {\bar{X}} \right)=\left( 1-{{P}_{D}} \right){{D}_{k|k-1}}\left( {\bar{X}} \right)+\sum\limits_{z\in {{\textbf{z}}_{k}}}{{{D}_{det ,k}}\left( \bar{X};z \right)},
\end{equation}
where
\begin{align}
\begin{split}\label{eq43}
{{D}_{det ,k}}\left( \bar{X};z \right)&=\sum\limits_{j=1}^{{{J}_{k|k\text{-}1}}\left( {{o}^{k}} \right)}{\omega _{k}^{j}\left( {{o}^{k}};z \right)}\\
&\times\mathcal{N}\left( \bar{X};\beta _{k|k-1}^{j}\left( {{o}^{k}} \right),m_{k}^{j}\left( {{o}^{k}} \right),U_{k}^{j}\left( {{o}^{k}} \right) \right),
\end{split}\\
\label{44}
\omega _{k}^{j}\left( {{o}^{k}};z \right)&=\frac{{{P}_{D}}\omega _{k|k-1}^{j}\left( {{o}^{k}} \right)q_{k}^{j}\left( {{o}^{k}};z \right)}{\kappa \left( z \right)+\sum\nolimits_{i=1}^{{{J}_{k|k-1}}}{\sum\nolimits_{{{o}^{k}}\in \mathbb{O}}{\omega _{k|k-1}^{i}\left( {{o}^{k}} \right)q_{k}^{i}\left( {{o}^{k}};z \right)}}},\\
\label{45}
q_{k}^{j}\left( {{o}^{k}};z \right)&=\left( z;Hm_{k|k-1}^{j,\left[ k \right]}\left( {{o}^{k}} \right),HP_{k|k-1}^{j,\left[ k \right]}\left( {{o}^{k}} \right){{H}^{\top}}+R \right),\\
\label{46}
m_{k}^{j}\left( {{o}^{k}} \right)&=m_{k|k-1}^{j}\left( {{o}^{k}} \right)\text{+}K\left( z-Hm_{k|k-1}^{j,\left[ k \right]}\left( {{o}^{k}} \right) \right),\\
\label{47}
U_{k}^{j}\left( {{o}^{k}} \right)&=U_{k|k-1}^{j}\left( {{o}^{k}} \right)-KHU_{k|k-1}^{j,\left[ k,: \right]}\left( {{o}^{k}} \right),\\
\label{48}
K&=U_{k|k-1}^{j,\left[ :,k \right]}\left( {{o}^{k}} \right){{H}^{\top}}{{\left( HU_{k|k-1}^{j,\left[ k \right]}\left( {{o}^{k}} \right){{H}^{\top}}+R \right)}^{-1}}.
\end{align}

Propositions 3 and 4 are the consequence of Propositions 1 and 2 and the properties of Gaussian density are shown by the Lemmas in~\cite{phd}. The recursion of the LGM-MM-TPHD filter is similar to the LGM-MM-PHD filter~\cite{mmphd}. Specially, the updated weights are the same as in the LGM-MM-PHD filter because the likelihood only depends on the current set of targets.
\subsection{L-Scan approximation}
Analogously to GMPHD filter, the number of Gaussian components for LGM-MM-TPHD filter increases as time progresses. Hence, to limit complexity, we still need to employ pruning and absorption techniques. The details of these techniques can be referred to~\cite{tcphd}.

In addition, the lengths of the trajectories increase with time, thereupon it is not computationally feasible to implement the proposed filters directly. To resolve this problem, the $L$-scan implementations that propagate the joint density of the states of the last $L$ time steps is presented. In the $L$-scan LGM-MM-TPHD filter, we discard the correlations of states that happened last $L$ time steps before the current time step in the prediction. Specifically, we adopt the independent Gaussian densities to represent the states outside the $L$-scan window and a joint Gaussian density for the states in the $L$-scan window. The implementation details is same as the $L$-scan GMTPHD filter, refer to~\cite{tcphd}.

\section{Simulation Results}

In this section, we demonstrate the performance of the proposed MM-TPHD filter with LGM implementation in a challenging multiple maneuvering targets tracking scenario which is referred to~\cite{mmvo}. The metric for trajectory RFS based on linear programming in~\cite{lptm} with parameters $p$ = $2$, $c$ = $10$ and $\gamma$ = $1$ is used to evaluate the performance.

Consider a two-dimensional surveillance area of $10000m\times 10000m$ with the duration is $K$ = $60s$ and a total of 5 maneuvering targets appeared during the duration. Targets 1, 2, 3, 4 and 5 enter the scene at times $k$ = $1,5,5,10,10s$ and targets 1, 2, 3 and 4 exit the scene at times $k$ = $40,40,50,50s$. Each target can randomly switch the motion model among three possible modes, where mode 1 is a CV model, mode 2 is a CT model with a counterclockwise turn rate of ${{10}^{\circ }}/s$ and mode 3 is also a CT model with a clockwise turn rate of ${{10}^{\circ }}/s$. The standard deviation of the process noise of the three modes is ${{\sigma }_{p}}=5m/{{s}^{2}}$. The linear state transition matrices for the CV and CT models as follows
\[{{F}_{CV}}={{I}_{2}}\otimes \left[ \begin{matrix}
1 & T  \\
0 & 1  \\
\end{matrix} \right],\]
\[{{F}_{CT}}=\left[ \begin{matrix}
1 & \sin \left( \theta T \right)/\theta  & 0 & -\left( 1-\cos \left( \theta T \right) \right)/\theta   \\
0 & \cos \left( \theta T \right) & 0 & -\sin \left( \theta T \right)  \\
0 & \left( 1-\cos \left( \theta T \right) \right)/\theta  & 1 & \sin \left( \theta T \right)/\theta   \\
0 & \sin \left( \theta T \right) & 0 & \cos \left( \theta T \right)  \\
\end{matrix} \right],\]
\[{{Q}_{CV}}={{Q}_{CT}}=\sigma _{p}^{2}{{I}_{2}}\otimes \left[ \begin{matrix}
{{T}^{4}}/4 & {{T}^{3}}/2  \\
{{T}^{3}}/2 & {{T}^{2}}  \\
\end{matrix} \right],\]
where $\otimes $ is the Kronecker product and $T$ = $1s$ is the sampling interval.
\begin{figure}
	\begin{center}
		\includegraphics[width=0.8\columnwidth,draft=false]{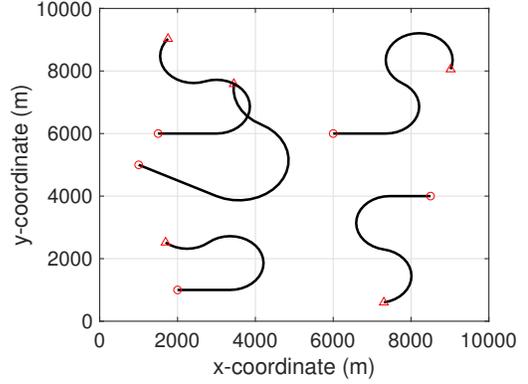}
	\end{center}
	\caption{The region and the trajectories in the ground truth: The start and end points for each trajectory are marked by $\bigcirc$ and $\bigtriangleup$, respectively.}
	\label{fig1}
\end{figure}
\begin{figure}
	\begin{center}
		\includegraphics[width=0.8\columnwidth,draft=false]{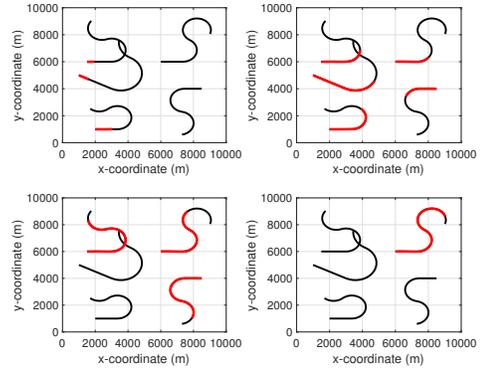}
	\end{center}
	\caption{Exemplar outputs at time steps 8 (upper left), 24 (upper right), 45 (lower left), 56 (lower right) of the MM-TPHD filter are shown in the surveillance area. The black lines represent the true trajectories. The red lines represent the estimated alive trajectories at current time.}
	\label{fig2}
\end{figure}

The Poisson birth intensity is a Gaussian mixture\eqref{eq31} with parameters: ${{J}_{\gamma ,k}}$ = $5$, $\omega _{\gamma ,k}^{j}\left( {{o}^{k}} \right)$ = $0.2p\left( {{o}^{k}} \right)$, $m_{\gamma ,k}^{1}$ = [2000; 0; 1000; 0], $m_{\gamma ,k}^{2}$ = [1000; 0; 5000; 0], $m_{\gamma ,k}^{3}$ = [1500; 0; 6000; 0], $m_{\gamma ,k}^{4}$ = [8500; 0; 4000; 0], $m_{\gamma ,k}^{5}$ = [6000; 0; 6000; 0] and $U_{\gamma ,k}^{j}$ = $diag{{\left( \left[ 10;10;10;10 \right] \right)}^{2}}$. The $p\left( o \right)$ is the model distribution at birth, which is taken as $p\left( o \right)$ = $\left[ 0.4,0.3,0.3 \right]$ and the switching between modes is given by the following Markovian model transition probability matrix (TPM):
\[\upsilon \left( o|{o}' \right)=\left[ \begin{matrix}
0.8 & 0.1 & 0.1  \\
0.1 & 0.8 & 0.1  \\
0.1 & 0.1 & 0.8  \\
\end{matrix} \right].\]

In addition, the standard deviations of the measurement noise is ${{\sigma }_{m}}$ = $10m$, and the clutter obeys Poisson distribution with clutter rate ${{\lambda }_{c}}$ = $60$. The survival probability and detection probability are ${{\text{P}}_{S}}$ = $0.99$, ${{\text{P}}_{D}}$ = $0.98$, respectively. The region and the trajectories of ground truths is presented in Fig. 1.

In the LGM implementation of MM-TPHD filter, we denote the pruning threshold as ${{\Gamma }_{p}}$ = ${{10}^{-5}}$, absorption threshold as ${{\Gamma }_{a}}$ = $4$ and limit the number of components to 30. Figure. 2 shows four exemplar outputs of the LGM-MM-TPHD filter, in which the filter provides an estimate of the set of present trajectories at the current time of each time step. Obviously, the MM-TPHD filter is capable to estimate the alive trajectories with high accuracy at each time step.

Then, we implement the $L$-scan approximation of the proposed filter with $L\in \left\{ 1,2,5 \right\}$ and use the metric for trajectory RFS to evaluate its performance by Monte Carlo simulation with 500 runs. The RMS trajectory errors for the $L$-scan MM-TPHD filter are shown in Fig. 3. As expected, increasing $L$ can improve estimation performance and reduce the errors as we take a longer time window to update the trajectories. The performance for $L\ge 5$  is similar to the MM-TPHD filter without $L$-scan approximation. Moreover, the single run time of Matlab implementation on the processor : Intel(R) Core(TM) i5-4590 CPU @ 3.30GHz, are approximately equal for $L\in \left\{ 1,2,5 \right\}$: $2.61$ seconds. However, if we proceed to increase the $L$, the single run time increases significantly, e.g, $5.34s$ for $L$ = $20$ and $10.52s$ for $L$ = $30$.

\begin{figure}
	\begin{center}
		\includegraphics[width=0.8\columnwidth,draft=false]{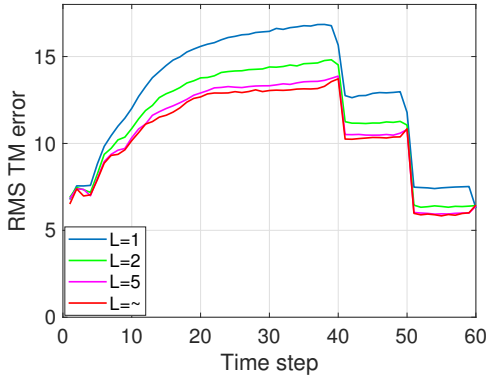}
	\end{center}
	\caption{The RMS trajectory metric error of the alive trajectories for the $L$-scan MM-TPHD filter. where $L$ = $\sim $ represents the error of MM-TPHD filter without $L$-scan approximation.}
	\label{fig3}
\end{figure}

\section{Conclusion}
The JMS model has proven to be an effective tool for multiple maneuvering target tracking who is a challenging research topic. A new algorithm based on TPHD filter for tracking the trajectories of multiple maneuvering targets is proposed with JMS model, named as MM-TPHD. The recursion of MM-TPHD filter is derived and the analytic closed-form is developed with linear Gaussian mixture implementation. To reduce computational burden, we present the \textit{L}-scan implementations of MM-TPHD for linear Gaussian model which is computationally efficient. We verify the trajectories tracking performance of the MM-TPHD filter via simulation results, based on the metric for trajectory RFS.







%




\end{document}